\documentclass[12pt]{article}
\input xy
\xyoption{all}
\xyoption{web}
\usepackage{amsmath}
\usepackage{amsfonts}
\usepackage{amssymb}

\newif\ifpdf
  \ifx \pdfoutput\undefined
    \pdffalse
  \else
    \pdfoutput=1 \pdftrue
  \fi
\ifpdf
  \usepackage[pdftex]{color,graphicx}
  \DeclareGraphicsExtensions{.pdf,.jpg,.png}
\else
  \usepackage{color,graphicx}
  \DeclareGraphicsExtensions{.ps,.eps}
\fi

\newlength{\abstractwidth}
\usepackage{epsf}

\renewcommand{\thefootnote}{\fnsymbol{footnote}}
\renewcommand{\thanks}[1]{\footnote{#1}}
\newcommand{\starttext}{
\setcounter{footnote}{0}
\renewcommand{\thefootnote}{\arabic{footnote}}}

\newcommand{\bea}{\begin{eqnarray}}
\newcommand{\eea}{\end{eqnarray}}
\newcommand{\ee}{\end{equation}}
\newcommand{\be}{\begin{equation}}

\def\no{\nonumber}

\begin{document}
\starttext
\setcounter{footnote}{0}

\begin{flushright}
UCLA/08/TEP/22
\end{flushright}

\bigskip

\begin{center}

{\Large \bf Instantons and Wormholes   for the Universal Hypermultiplet}

\vskip .7in 

{\large Marco Chiodaroli and Michael Gutperle}

\vskip .2in

 \sl Department of Physics and Astronomy \\
\sl University of California, Los Angeles, CA 90095, USA

\end{center}

\vskip .5in

\begin{abstract}

In this paper we present supersymmetric instanton and non-supersymmetric wormhole solutions for the universal hypermultiplet sector of $d=4$ $N=2$ supergravity theories.
Instantons and wormholes are constructed as saddle points dominating transition amplitudes between states of definite axionic shift charge, using an approach due to Coleman and Lee.
Our solutions are  constructed
in terms of the conserved Noether charges associated with the global $SU(2,1)$ symmetry of the universal hypermultiplet action.
The conditions imposed by regularity on the charges are discussed.
 
\end{abstract}

\newpage

 \baselineskip=18pt
\setcounter{equation}{0}
\setcounter{footnote}{0}

\section{Introduction}

 In string theory, instantons and wormholes can be constructed as extrema of the low energy supergravity Euclidean action \cite{Hawking:1988ae,Lavrelashvili:1988jj,Giddings:1987cg,Coleman:1989zu}.  
 
It has been argued in the past that wormholes lead to bi-local terms in the low energy effective action and produce several interesting effects, such as renormalization of coupling constants and cosmological constant, quantum decoherence and creation of baby universes \cite{Lavrelashvili:1988jj,Giddings:1987cg,Strominger:1983ns,Hawking:1987mz,Giddings:1988wv,Klebanov:1988eh,Coleman:1988cy}. 
Wormholes in Anti de-Sitter spaces have been investigated in \cite{Rey:1998yx,Gutperle:2002km,Maldacena:2004rf},
while some recent work \cite{ArkaniHamed:2007js} has pointed out how wormhole-induced effects determine a clash between the locality of an Anti de-Sitter theory and the locality of the dual conformal field theory.
 
Instantons, on the other hand, lead to local non-perturbative contributions to the effective action. In supergravity theories, the BPS instanton solutions of the Euclidean equations of motion preserve half of the supersymmetries \cite{Gibbons:1995vg,Green:1997tv}. They can be viewed as an extremal limit of non-supersymmetric wormhole solutions,  where the  neck of the wormhole  pinches off, leading to a local operator insertion. 

The broken supersymmetries of the instanton solution generate fermionic zero modes. Transition amplitudes will vanish, unless the zero modes are soaked up by the insertion of appropriate operators in the path integral \cite{Green:1997tv,Becker:1995kb}.  This leads to instanton-induced terms in the effective action,  
including four-fermion terms as well as corrections to the sigma model metric of the universal hypermultiplet.
  Instanton and wormhole solutions have been discussed for various theories and dimensions, in particular for the axion/dilaton $SL(2,R)/U(1)$ coset \cite{Giddings:1989bq,Gibbons:1995vg,Green:1997tv,Bergshoeff:2004fq,Einhorn:2002sj}, the universal hypermultiplet in $N=2,d=4$ supergravity \cite{Behrndt:1997ch,Gutperle:2000sb,Davidse:2003ww,Davidse:2004gg,Theis:2002er,Ketov:2003hc,Bergshoeff:2004pg} and general hypermultiplets in $N=2,d=4$ theories \cite{Gutperle:2000ve,deVroome:2006xu,Bergshoeff:2008be}.

In this paper we focus on instanton and wormhole solutions for the universal hypermultiplet sector of $N=2$ $d=4$ supergravities with vanishing cosmological constant.
The universal hypermultiplet always appears as a subsector of the hypermultiplet sector of Calabi-Yau compactifications of type II string theories. In the case of type IIA on a so-called rigid Calabi-Yau (which has $h_{2,1}=0$), the universal hypermultiplet constitutes the complete hypermultiplet sector. 

The main motivation of this paper is to apply the approach originally due to Coleman and Lee \cite{Coleman:1989zu} to the universal hypermultiplet. In section 2 we review the universal hypermultiplet and its symmetries.  In section 3 we generalize the approach from Coleman and Lee to the case of non-commuting shift charges. Instanton and wormhole amplitudes are constructed in the path integral framework as transition amplitudes between states of definite axionic shift charges. This allows us to obtain positive-definite Euclidean  actions and clarifies the procedure of analytic continuation for axionic scalars. 
In section 4 we construct the general instanton and wormhole solutions in terms of the conserved charges derived in section 2. The extremal (BPS) limit is discussed in section 5. The integrability of the supersymmetry variations is verified and the instanton action is calculated. In section 6 we discuss the constraints on non-extremal wormhole solutions imposed by regularity. In three appendices we present some detailed formulae
for the NS-R charged solution, the relation of our solution to the one obtained by dualizing axionic scalars to tensor fields and the conserved charges after analytic continuation.

\section{The universal hypermultiplet}
The universal hypermultiplet action contains four scalar fields denoted with $\phi,\sigma,\zeta,\tilde \zeta$. The parameterization which is most useful for our purposes is given by:
\begin{equation}
S= \int d^{4}x \sqrt{g} \;\Big\{ {1\over 2} (\partial_{\mu} \phi)^{2} +{1\over 2}e^{-2\phi}(\partial_{\mu} \sigma + \tilde \zeta \partial_{\mu} \zeta)^{2}+{1\over 2}e^{-\phi}\big[ (\partial_{\mu} \zeta)^{2}+(\partial_{\mu }\tilde\zeta)^{2}\big]\Big\}
\label{unione}
\end{equation}
 
This action can be obtained by compactification of type IIA string theory on a rigid Calabi-Yau manifold with $h_{2,1}=0$.  In all Calabi-Yau compactifications, it is possible to find a consistent truncation reducing the hypermultiplet action to one equivalent to the universal hypermultiplet. 

The action (\ref{unione}) can also be written as a gauge-fixed sigma model action for the $\frac{SU(2,1)}{SU(2)\times U(1)}$ coset. The coset possesses a nonlinearly realized  global $SU(2,1)$ symmetry. The eight infinitesimal generators $\Phi^{i}\to \Phi^{i}+ \epsilon  \;\delta_{a} \Phi^{i}$ are given\footnote{We use the same labeling of generators as  in  \cite{Gunaydin:2007qq}.} as follows. 
First, the shift symmetry of the NS-NS axion $\sigma$ is generated by $E$, the simple shift of the R-R axion $\zeta$ is generated by $E_{q}$ and the shift of the second R-R axion $\tilde \zeta$ is generated by $E_{p}$:
\begin{eqnarray} \label{shifta}
E:&&  \quad  \delta \sigma = 1    \\
E_{q}:&& \quad \delta \zeta = - \sqrt{2}   \\
 E_{p}:&&\quad \delta \tilde \zeta= - \sqrt{2}, \quad \delta \sigma = \sqrt{2} \zeta
\end{eqnarray}
Second, a scaling symmetry is generated by $H$ and a rotation of $\zeta$ and $\tilde \zeta$ is generated by $J$:
\begin{eqnarray}
 H:&& \delta \phi = 2 , \quad \delta \sigma = 2\sigma, \quad \delta \zeta = \zeta, \quad \delta \tilde\zeta= \tilde \zeta  \\
J:&& \delta \phi = 0 , \quad \delta \sigma = {1\over 2}(\tilde\zeta^{2}-\zeta^{2}) , \quad \delta \zeta =- \tilde \zeta, \quad \delta \tilde\zeta=   \zeta 
\end{eqnarray}
Third, there are three  additional global symmetries which complete the $SU(2,1)$ algebra:
\begin{eqnarray}
\no F_{p}:&& \delta \phi =\sqrt{2} \tilde \zeta, \quad  \delta \sigma = {\sqrt{2}\over 4}( \zeta^{3}- 3\zeta \tilde \zeta^{2}) \\
&&  \delta\zeta=\sqrt{2} \left( \sigma +{3\over 2} \zeta \tilde \zeta \right), \quad \delta \tilde \zeta= - \sqrt{2} \left( e^{\phi}+{3\over 4}  \zeta^{2}-{1\over 4}\tilde \zeta^{2} \right)  \\
\no F_{q}:&& \delta \phi =\sqrt{2} \zeta, \quad  \delta \sigma = \sqrt{2} \left( \sigma\zeta + e^{\phi}\tilde\zeta +{1\over 2}\tilde \zeta^{3} \right), \\
&& \delta\zeta=\sqrt{2}\left(- e^{\phi}+{1\over 4} \zeta^{2}-{3\over 4} \tilde\zeta^{2} \right), \quad \delta \tilde \zeta= \sqrt{2} \left(- \sigma +{1\over 2} \zeta\tilde\zeta \right)  \\
\no F:&& \delta \phi = - (2\sigma+\zeta\tilde \zeta), \quad \delta\sigma = e^{2\phi}-\sigma^{2 } + e^{\phi }\tilde \zeta^{2}-{1\over 16} \zeta^{4}+{3\over 16} \tilde\zeta^{4}+{3\over 8} \zeta^{2}\tilde \zeta^{2} \\
 && \delta \zeta= -\sigma \zeta-e^{\phi}\tilde \zeta -{3\over 4}\zeta^{2}\tilde\zeta-{1\over 4} \tilde\zeta^{3},
  \quad \delta\tilde\zeta= - \sigma \tilde \zeta+e^{\phi}\zeta +{1\over 4}\zeta^{3}-{1\over 4} \zeta \tilde\zeta^{2}
\end{eqnarray}
The eight global symmetries lead to eight  Noether currents given by:
\begin{equation}
j^{\mu}_{a} = \sum_{i=1}^{4}{\delta  L\over \delta(\partial_{\mu } \Phi_{i})} \delta_{{a}} \Phi_{i}
\end{equation}
\begin{figure}[htbp]
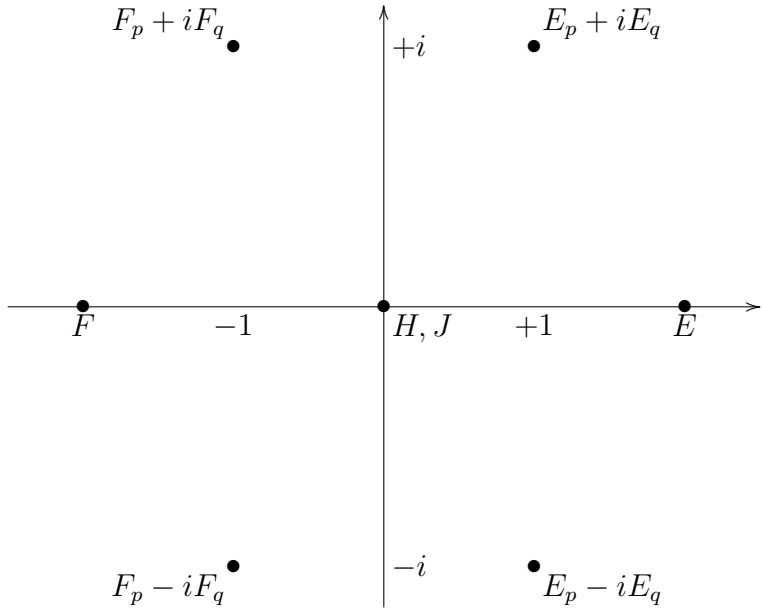

\begin{center}
\xy
\POS (0,0) *{\bullet}*+!UL{H, J}="HJ"
,(20,34.6) *{\bullet}*+!DL{E_p+i E_q}="E+"
,(20,-34.6) *{\bullet}*+!UL{E_p-i E_q}="E-"
,(-20,34.6) *{\bullet}*+!DR{F_p+i F_q}="F+"
,(-20,-34.6) *{\bullet}*+!UR{F_p-i F_q}="F-"
,(40,0) *{\bullet}*+!U{E}="E"
,(-40,0) *{\bullet}*+!U{F}="F"
,(-80,0) ="xxL"
,(-50,0) ="xL"
,(50,0) ="xR"
,(70,0) ="xxR"
,(0,40) = "yU"
,(0,-40) = "yD" 
,(-20,0) *+!U{-1}
,(20,0) *+!U{+1}
,(0,34.6) *+!L{+i}
,(0,-34.6) *+!L{-i}
\POS "xxL" \ar@{} "xL"
\POS "xL" \ar@{->} "xR"
\POS "yD" \ar@{->} "yU"
\endxy 
 \caption{Root diagram of $SU(2,1)$ with identification of the symmetry generators.}
\end{center}
\end{figure}
where the index $i$ runs over the hypermultiplet fields and $a\in \{E,E_{p},E_{q},H,J,F,F_{p},F_{q}\}$ labels the symmetry generator. On shell, the eight currents are conserved.
\begin{equation}
\partial_{\mu}\big( \sqrt{g} g^{\mu\nu}j_{a \nu} \Big)=0, \quad a\in \{E,E_{p},E_{q},H,J,F,F_{p},F_{q}\}
\end{equation}
The supersymmetry transformation rules  of  hypermultiplets coupled to the $N=2$ supergravity multiplet were  found in \cite{Bagger:1983tt} and will be reviewed in section 
\ref{secsusy}.
 
\section{Coleman's approach}
\setcounter{equation}{0}
Instantons and wormholes solutions are saddle points of the Euclidean action.
Regular instanton and wormhole solutions in  theories which contain axionic scalars only exist if the sign of the kinetic term for the axionic scalar is flipped as the theory is continued from Minkowski to Euclidean spacetime. This procedure appears at first sight  to be somewhat ad-hoc and potentially ambiguous. 
There are several approaches to deal with this issue. One is to dualize axions to rank three  antisymmetric tensor fieldstrengths \cite{Giddings:1987cg,Green:1997tv} and rewrite the universal hypermultiplet in terms of a tensor multiplet action
 \cite {Theis:2002er} where the Wick rotation poses no problems.  A more formal approach is to replace the Minkowskian quaternionic geometry by a para-quaternionic geometry in Euclidean space \cite{Gibbons:1995vg,Cortes:2005uq}. Here we apply a third approach originally due to Coleman and Lee \cite{Coleman:1989zu} to the universal hypermultiplet (see \cite{ArkaniHamed:2007js} for a recent  application for the axion in the $SL(2,R)/U(1)$ coset).
 
One considers imaginary-time transition amplitudes between initial and final states with constant values of the hypermultiplet fields:
\bea
| I \rangle &=& |  \phi_I ,\zeta_I, \tilde{\zeta}_{I},\sigma_I \rangle \nonumber \\ 
| F \rangle &=& | \phi_F, \zeta_F, \tilde{\zeta}_{F},\sigma_F \rangle
\eea
 We can project initial and final states into eigenspaces of the shift charge densities. For the universal hypermultiplet, there is however a complication due to the fact that the charge densities of the  three shift symmetries (\ref{shifta}) obey a nontrivial commutation relation
\begin{equation}
\big[ j^{\tau }_{E_{q}}(\vec{x},\tau), j^{\tau}_{E_{p}}(\vec{y}, \tau)\big]= 2\delta^{3}(\vec{x}-\vec{y})j^{\tau}_{E }(\vec{x},\tau) \label{comm}
\end{equation}
 which defines a  Heisenberg algebra. Consequently the initial and final eigenstates cannot be projected on charge eigenstates of all the three shift charge densities $j^{\tau}_{E},j^{\tau}_{E_{q}},j^{\tau}_{E_{p}}$.

According to the Stone-von Neumann theorem, a representation of the Heisenberg algebra is uniquely labelled by the value of the charge associated with the central element $E$. We have to distinguish two cases depending on whether the central element vanishes or not.
In case of vanishing $E$ charge density, the commutation relation (\ref{comm}) becomes trivial and we can project initial and final states into charge density eigenspaces of both $E_p$ and $E_q$. The imaginary-time transition amplitude is dominated by saddle points charged only under the shifts of the R-R scalars ("pure R-R charged" instantons and wormholes).

In case of non vanishing $E$ charge density, the best we can do is to project initial and final states into charge density eigenspaces of $E_q$ and $E$. The relevant saddle points will be charged also under the shift of the NS scalar $\sigma$ ("mixed NS-R charged" instantons and wormholes). 

 \subsection{Pure R-R charged case}

For a zero value of the charge density $j_{E}$, the initial and final states can be projected onto eigenstates of $j_{E_{p}}$ and $j_{E_{q}}$.
 The projection for the initial state acts as follows: 
 \begin{eqnarray}
 P_{I}\mid I\rangle &=& \delta(j^{\tau}_{E}) \delta(j^\tau_{E_{p}}-\rho^{I}_{E_{p}}) \delta(j^\tau_{E_{q}}-\rho^{I}_{E_{q}})|I\rangle \nonumber\\ 
 &\propto &\int \mathcal{D}\alpha  \mathcal{D}\gamma  \mathcal{D}\tilde \gamma  \;e^{i \int d^3 \vec{x} \; \alpha \; j^{\tau}_{E}} e^{i \int d^3 \vec{x}\;  \tilde \gamma(j^\tau_{E_{p} }-\rho^{I}_{E_{p}})}e^{i \int d^3 \vec{x}\;  \gamma(j^\tau_{E_{q} }-\rho^{I}_{E_{q}})}
 \mid \phi_I,\zeta_I , \tilde{\zeta}_{I }, \sigma_I \rangle \nonumber \\
 &=& \int \mathcal{D}\alpha \mathcal{D}\gamma \mathcal{D}\tilde \gamma e^{-i \int d^3 \vec{x} (\tilde \gamma \rho^{I}_{E_{p}} + \gamma \rho^{I}_{E_{q}} )} | \phi_I, \zeta_{I}- \sqrt{2}  \gamma, \tilde\zeta_{I}-\sqrt{2} \tilde \gamma , \sigma_{I}+\alpha+ \sqrt{2} \tilde\gamma \zeta_{I} \rangle \end{eqnarray}
The final state projection works analogously. The three-dimensional integrals are over the constant Euclidean time surfaces.
Redefining:
\begin{equation}
\zeta = \zeta^{I} - \sqrt{2}\gamma,\quad \tilde \zeta = \tilde \zeta^{I} - \sqrt{2}\tilde \gamma, \quad \sigma = \sigma^{I}+\alpha + \sqrt{2}\tilde\gamma\zeta^{I}\label{redefrr}
\end{equation}
The transition amplitude becomes:
\be
\langle F | P_{F }e^{-H (\tau_F- \tau_I)}  P_{I}| I \rangle =
e^{i \int d^{3}\vec{x} \frac{1}{\sqrt{2}}(\rho_{E_{q}}^{F} \zeta_F -\rho_{E_{q}}^{ I}\zeta _I + \rho_{E_{p}}^{F} \tilde\zeta_F- \rho_{E_{p}}^{ I}\tilde\zeta _I ) } \int \mathcal{D}\phi   \mathcal{D}\zeta  \mathcal{D}\tilde \zeta  \mathcal{D}\sigma e^{-(S_E+\Sigma)} \label{transamp}
\ee
The phase factor on the right hand side of  (\ref{transamp}) depends on the initial and final values of the original axions as well as the charges. $\Sigma$ is a surface term given by:
\be
\Sigma=i \int{d^3\vec{x} \frac{1}{\sqrt{2}} \left[\rho_{E_{p}}^{F}(\vec{x}) \tilde \zeta(\vec{x},\tau_F)-\rho_{E_{p}}^{I}(\vec{x}) \tilde\zeta  (\vec{x},\tau_I)+ \rho_{E_{q}}^{F}(\vec{x}) \zeta  (\vec{x},\tau_F)-\rho_{E_{q}}^{I}(\vec{x}) \zeta(\vec{x},\tau_I)\right]}
\ee

As an effect of the change of variables  (\ref{redefrr}), the functional integration of the redefined fields $\sigma, \zeta, \tilde \zeta$ goes over configurations without fixed initial and final values. In particular $\zeta(\vec{x},\tau_{I,F})$ and $\tilde{\zeta}(\vec{x},\tau_{I,F})$ do not equal the initial or final values  $\zeta_{I,F}(\vec{x})$ and $\tilde{\zeta}_{I,F}(\vec{x})$ respectively. \\
As a consequence, the variational principle determining the saddle point of the path integral contains nontrivial boundary contributions from the initial and final time. The boundary term from the variation of $S_{E}+\Sigma$ is given by
\be
\int d^{3}\vec{x} \left. \left[ \big(i \rho_{E_{p}}^{F} - j^{\tau}_{E_{p}} \big) \frac{ \textstyle \delta \tilde\zeta}{\sqrt{2}} + \big( i \rho_{E_{q}}^{F} - j^{\tau}_{E_{q}} \big) \frac{ \delta \zeta}{\sqrt{2}} \right] \right|_{\tau_{F}}- \int d^{3}\vec{x} \left. \left[ \big(i \rho_{E_{p}}^{I} - j^{\tau}_{E_{p}} \big)\frac{ \delta \tilde\zeta}{\sqrt{2}}+  \big( i \rho_{E_{q}}^{I} - j^{\tau}_{E_{q}} \big)\frac{ \delta \zeta}{\sqrt{2}}  \right] \right|_{\tau_{I}} \nonumber
\ee
The vanishing of the boundary term  imposes boundary conditions on the fields $\zeta$ and $\tilde \zeta$ which cannot be satisfied by real fields. Hence  the path integral is dominated by a complex saddle point where both $\zeta$ and $\tilde \zeta$ are pure imaginary. We can make the saddle point real by the analytic continuation 
 \be
\zeta \to  i \zeta' ,\quad  \tilde \zeta \to  i \tilde \zeta' \label{anaconrr}
\ee
The analytic continuation of the fields causes an analogous analytic continuation for the conserved Noether currents. 
\be j^\tau_{E_q} \rightarrow i \; {j'^\tau_{E_q}}, \quad j^\tau_{E_p} \rightarrow i\; j'^\tau_{E_p} \ee
An expression of the continued currents in terms of the primed fields is given in the appendix. This analytic continuation destroys the positive-definiteness of the bulk Euclidean action $S_{E}$ since the sign of the kinetic terms for $\zeta'$ and $\tilde \zeta'$ becomes negative. 
Using the equations of motion and the analytic continuation (\ref{anaconrr}), the surface term can be rewritten as \footnote{We use the fact that for $SO(4)$ invariant solutions of the equations of motion  $j'^{\tau}_{E}=0$ at one point implies $j'^{\mu}_E=0$ everywhere. This simplifies the form of the Noether currents $j'_{E_{q}}$ and $j'_{E_{p}}$.}:
\bea
 \Sigma &=& - \frac{1}{\sqrt{2}} \int d^4x \;\partial_{\mu} \big[\sqrt{g}  (j'^{\mu}_{E_{p}} \tilde \zeta'+j'^{\mu}_{E_{q}} \zeta') \big]\nonumber\\
 &=& - \frac{1}{\sqrt{2}} \int d^4x \sqrt{g}\big(  j'^{\mu}_{E_{p}} \; \partial_\mu  \tilde \zeta' +  j'^{\mu}_{E_{q}} \; \partial_\mu  \zeta' \big)\nonumber\\
 & =&\int{d^4x \sqrt{g} \; e^{\phi'}\big( \partial_{\mu} \zeta' \partial^{\mu }\zeta' +\partial_{\mu }\tilde\zeta' \partial^{\mu }\tilde \zeta'\big) } 
\eea
 The total saddle point action $S_{E}+\Sigma$  is manifestly positive-definite:
\be
S_E+\Sigma = \int d^4x \sqrt{g}\; \Big[-R +  \frac{1}{2} \partial_\mu \phi'\partial^{\mu}\phi'    +  {1\over 2} e^{\phi'}\big( \partial_{\mu} \zeta' \partial^{\mu }\zeta' +\partial_{\mu }\tilde\zeta' \partial^{\mu }\tilde \zeta' \big)\Big]
 \ee

 \subsection{Mixed NS-R charged case}
 For a non-zero value of the charge density $\rho_{E}$, the initial and final states are projected onto eigenstates of fixed $\rho_{E}$ and $\rho_{E_q}$. Following similar steps as in the pure R-R case one gets:
\bea
 P_{I}|I\rangle&=& \delta(j_{E}^{\tau}-\rho_{E}^{I})  \delta(j_{E_{q}} -\rho_{E_{q}}^{I})  |\phi_{I}
  \zeta_I , \tilde \zeta_{I}, \sigma_{I}\rangle \nonumber  \\
  &\propto& \int \mathcal{D}\alpha  \mathcal{D}\gamma  \;e^{i \int d^3 \vec{x} \; \alpha (j^{\tau}_{E}-\rho_{E}^{I})}  \; e^{i \int d^3 \vec{x}\;  \gamma(j^\tau_{E_{q} }-\rho^{I}_{E_{q}})} \mid \phi_I,\zeta_I , \tilde{\zeta}_{I }, \sigma_I \rangle
\nonumber  \\ 
&=& \int{\mathcal{D}\alpha  \mathcal{D}\gamma \; e^{-i \int{d^3 \vec{x} (\alpha \rho^I_{E} + \gamma \rho_{E_{q}}^{I} )}}| \phi_I, \zeta_{I} - \sqrt{2}\gamma, \tilde \zeta_I, \sigma_{I}+\alpha \rangle }
\eea  
Redefining the scalar fields:
\be
\quad \zeta=\zeta_I- \sqrt{2} \gamma,  \quad \tilde\zeta=\tilde \zeta_I,  \quad \sigma=\sigma_I+\alpha
\ee
The transition amplitude can be written as a path integral over $\zeta$ and $\tilde \zeta$: 
\be \langle F | P_{F }e^{-H (\tau_F- \tau_I)}  P_{I}| I \rangle= 
e^{i \int d^{3}\vec{x} (\rho_{E}^{ I}\sigma _I -\rho_{E}^{F} \sigma_F -\frac{1}{\sqrt{2}}\rho_{E_{q}}^{ I} \zeta _I + \frac{1}{\sqrt{2}}\rho_{E_{q}}^{F} \zeta_F) } \int \mathcal{D}\phi   \mathcal{D}\zeta  \mathcal{D}\tilde \zeta  \mathcal{D}\sigma e^{-(S_E+\Sigma)} \ee 
The surface term $\Sigma$ is given by:
\be
\Sigma=i \int{d^3 \vec{x}\; \Big( \rho^{I}_{E}(\vec{x}) \sigma(\vec{x},\tau_I)- \rho^{F}_{E}(\vec{x})\sigma(\vec{x},\tau_F)+  \rho^{F}_{E_{q}} (\vec{x}) \frac{\zeta (\vec{x},\tau_F)}{\sqrt{2}} - \rho^{I}_{E_{q}} (\vec{x}) \frac{\zeta (\vec{x},\tau_I)}{\sqrt{2}}  \Big)}
\ee

 As in the pure R-R charged case, the variational principle of $S_{E}+\Sigma$ determining the saddle point of the path integral over the primed fields has a nontrivial  boundary term 
\be
- \int d^{3}\vec{x} \left. \Big[ \big(i \rho_{E}^{F} - j^{\tau}_{E} \big) \delta \sigma -\big( i \rho_{E_{q}}^{F} - j^{\tau}_{E_{q}} \big) \frac{ \delta \zeta}{\sqrt{2}} -  \Big] \right|_{\tau_{F}} + \int d^{3}\vec{x} \left. \Big[\big(i \rho_{E}^{I} - j^{\tau}_{E} \big) \delta \sigma-\big( i \rho_{E_{q}}^{I} - j^{\tau}_{E_{q}} \big) \frac{\delta \zeta}{\sqrt{2}}  \Big] \right|_{\tau_{I}} \nonumber
\ee
The path integral is dominated by a complex saddle point where both $\zeta'$ and $\sigma'$ are pure imaginary. We can make the saddle point real by the analytic continuation
\be
\zeta \to i \zeta' ,\quad \sigma \to i \sigma' \label{anaconnsr}
\ee
The Noether currents after the analytic continuation are given in the appendix. Using the equations of motion the surface term can be rewritten as follows:
\bea 
\Sigma &=& \int d^4x \; \partial_{\mu} \sqrt{g} \left( j'^{\mu}_{E} \sigma'-\frac{1}{\sqrt{2}} j'^{\mu}_{E_{q}}  \zeta'   \right) =  \int d^4x \sqrt{g}\left( j'^{\mu}_{E} \partial_\mu {\sigma}' -\frac{1}{\sqrt{2}} j'^{\mu}_{E_{q}} \partial_\mu \zeta'  \right) \nonumber \\
&=& \int d^4x \sqrt{g} \Big [ e^{2\phi'}(\partial_{\mu} \sigma' + \tilde \zeta' \partial_{\mu} \zeta')^{2}+e^{\phi'} (\partial_{\mu} \zeta')^{2}\Big]
\eea
The bulk part of the action (\ref{unione}) after the analytic continuation (\ref{anaconnsr}) is not positive-definite and can be written as:
\begin{equation}
S_{E}= \int d^{4}x \sqrt{g} \;\Big\{ -R+ {1\over 2} (\partial_{\mu} \phi')^{2} -{1\over 2}e^{2\phi'}(\partial_{\mu} \sigma' + \tilde \zeta' \partial_{\mu} \zeta')^{2}+{1\over 2}e^{\phi'}\big[ -(\partial_{\mu} \zeta')^{2}+(\partial_{\mu }\tilde\zeta')^{2}\big]\Big\}
\label{unitwo}
\end{equation}
If we take into account the surface terms, the total action is positive definite:
 \begin{equation}
S_{E}+\Sigma= \int d^{4}x \sqrt{g} \;\Big\{ -R + {1\over 2} (\partial_{\mu} \phi')^{2} +{1\over 2}e^{2\phi'}(\partial_{\mu} \sigma' + \tilde \zeta' \partial_{\mu} \zeta')^{2}+{1\over 2}e^{\phi'}\big[ (\partial_{\mu} \zeta')^{2}+(\partial_{\mu }\tilde\zeta')^{2}\big]\Big\}
\label{unitwo}
\end{equation}
  
   \section{Extremal and nonextremal solutions}
   \setcounter{equation}{0}
In this paper we consider only $SO(4)$ invariant solutions of the equations of motion obtained from varying the Euclidean action. We use the following  ansatz for the Euclidean metric:
\be
ds_E^2={e^{3U}\over 4 \tau^3}d \tau^2+ {e^U\over \tau}d\Omega_{S_{3}}^2\label{metansatz}
\ee
Here $d \Omega^2_{S_{3}}$ is the metric of the three-sphere.  The ansatz reduces to the flat metric in case $U(\tau)\equiv 0$, with the identification $\tau=1/r^2$. Moreover, as a consequence of the $SO(4)$ invariance, all hypermultiplet  scalars  depend only on $\tau$.
With this choice for the metric one obtains from  Einstein  equation :
\begin{eqnarray}
  G_{\tau \tau}&= & \frac{3}{4 \tau^2} (1- e^{2U}-2 \tau \dot U +\tau^2 \dot U^{ 2})= T_{\tau \tau}   \label{einsteina}\\
G_{\alpha \beta }&=& e^{-2U} \big(1-e^{2U}+6 \tau \dot U - 3 \tau^2 \dot U^{ 2} + 4 \tau^2 \ddot U \big)\eta_{\alpha\beta}=T_{\alpha\beta}  \label{einsteinb}
\end{eqnarray}
where $\eta_{\alpha\beta}$ is the metric on the unit three-sphere and we denoted derivatives with respect to $\tau$ by dots. The energy-momentum tensor obeys to:
\begin{equation}
T_{\alpha\beta}= -4 \tau^2 e^{-2U} T_{\tau \tau}\;  \eta_{\alpha\beta}
\end{equation}
A linear combination of (\ref{einsteina}) and (\ref{einsteinb}) does not depend on the energy momentum tensor and gives the second order differential equation:
\be \ddot U =\frac{e^{2U}-1}{\tau^2} \label{Uode} \ee
All solutions can be brought in a form where $U(\tau)\rightarrow 0$ as $\tau \rightarrow 0$ with a simple rescaling of the radial coordinate $\tau$. Then, the general  solution of (\ref{Uode}) has the form:
\be e^{U(\tau)}=\frac{4\sqrt{c}\;  \tau}{\sin{( 4\sqrt{c} \; \tau})} \label{WHSol} \ee  
The radial coordinate $\tau$ can assume any value from $0$ to $\frac{\pi}{4 \sqrt{c}}$. These solutions are regular and exhibit two flat asymptotic regions ($\tau \rightarrow 0$ and $\tau \rightarrow \frac{\pi}{4 \sqrt{c}}$) connected by a wormhole  if $c> 0$. 
If $c<0$ the solution of the (\ref{Uode}) can be rewritten as:
\be e^{U(\tau)}=\frac{4 \sqrt{|c|}\;  \tau}{\sinh{(4 \sqrt{|c|} \tau)}} \label{SingSol}\ee
These solutions are singular for $\tau \rightarrow \infty$ and will not be studied in this paper. \\
Taking the limit $c\to 0$ for the wormhole solution (\ref{WHSol}) gives $U=0$ and hence one obtains  the flat metric. Solutions of this kind are the extremal instantons. \\

Furthermore, the $\tau \tau$ component of the Einstein equation produces a constraint. For the pure R-R-charged case one gets:
\begin{equation}
- 24\;  c= {1\over 2} (\partial_{\tau} \phi')^{2} -{1\over 2}e^{\phi'}\big( (\partial_{\tau} \zeta')^{2}+(\partial_{\tau }\tilde\zeta')^{2}\big) \label{consta}
\end{equation}
whereas for the mixed NS-R charged case one has:
\begin{equation}
-24 \;c= {1\over 2} (\partial_{\tau} \phi')^{2} -{1\over 2}e^{2\phi'}(\partial_{\tau} \sigma' +\tilde \zeta' \partial_{\tau} \zeta')^{2}-{1\over 2}e^{\phi'}\big( (\partial_{\tau} \zeta')^{2}-(\partial_{\tau }\tilde\zeta')^{2}\big) \label{constb}
\end{equation}
It follows from Einstein equation (\ref{einsteina}) and (\ref{einsteinb}) that the on-shell action vanishes for all values of $c$. Hence the action of the wormhole is given by the boundary term $\Sigma$ alone.

\section{Solution in terms of charges}
\setcounter{equation}{0}

On shell, the Noether currents are conserved ($D_{\mu} j^{\mu}_{{a}}=0$).   Using an $SO(4)$ invariant Euclidean solution and the metric (\ref{metansatz}) the conservation condition simply becomes $\partial_{\tau}j^{\tau}_{ a}=0$ and hence the currents are all constant.
 
 \begin{equation}
j^{\tau}_{{a}} = q_{{a}}, \quad  j_{a}^{\alpha} = 0 , \quad \quad a\in \{E,E_{p},E_{q},H,J,F,F_{p},F_{q}\}, \quad \alpha\in\{\theta , \phi, \psi \} 
\end{equation}
The conservation equations are equivalent to the equations of motion and it is possible to reduce the problem to a single first order differential equation for the dilaton. \\
Firstly, one uses the shift charges $q_{E},q_{E_{p}},q_{E_{q}}$ to solve for $\sigma,\zeta$ and $\tilde \zeta$. Then, the appropriate linear combinations of the equations for the charges $q_{H},q_{Fp},q_{Fq}$ can be used to solve for the scalars $\sigma, \zeta, \tilde \zeta$ in terms of $\phi$. \\
 The equation for $q_{F}$ gives a first order differential equation for $\phi$ and the equation for $q_{J}$ gives a cubic constraint in the charges.
  The solution can be expressed in terms of the charges and one additional integration constant. 

 \subsection{Pure R-R charged case}

 In the pure R-R charged case, the projection on $j_{E}^{\tau}=0$ implies that the charge $q_{E}$ vanishes.\\
 After the analytic continuation (\ref{anaconrr})  the metric parameter $c$ is related to the quadratic invariant in the charges via the constraint (\ref{consta}).
  \begin{equation}\label{metparb}
 c  =  {1\over 96}\Big(  q'_{E_{q}} q'_{F_{q}} + q'_{E_{p}}q'_{F_{p}} \Big) - {1\over 192} q'^2_{h} + {1\over 64} q'^2_{J}
\end{equation}
The differential equation for $\phi'$ simplifies if it is expressed in terms of the metric parameter $c$ using (\ref{metpar}).
\begin{equation}\label{dileqc}
(\partial_\tau \phi')^{2} -\frac{1}{2} e^{\phi'}\big( q'^2_{E_{q}}+ q'^2_{E_{p}} \big)
+ 48 c=0
\end{equation}
The cubic constraint is:
\begin{equation}
q'^3_{J}+ q'_{J} \big( q'^2_{H}+ q'_{E_{q}}q'_{F_{q}} + q'_{E_{p}}q'_{F_{p}}\big) + q'_H(q'_{E_{q}}q'_{F_{p}}  - q'_{E_{p}}q'_{F_{q}}) -  q'_F (q'^2_{E_{q}}  + q'^2_{E_{p}})=0\end{equation}
The R-R scalars are given by:
 \begin{eqnarray}
 \zeta' &=& {\sqrt{2}\over q'^2_{E_{p}}+q'^2_{E_{q}}}\big( q'_{E_{q}}q'_{H}+ q'_{E_{p}}q'_{J}- 2 q'_{E_{q}} \partial_\tau \phi'\big)\label{zetaeqa}\\
 \tilde\zeta' &=& {\sqrt{2}\over q'^2_{E_{p}}+q'^2_{E_{q}}}\big(q'_{E_{p}}q'_{H}- q'_{E_{q}}q'_{J}- 2 q'_{E_{p}}\partial_\tau \phi'\big)\label{zetaeqb}
\end{eqnarray}
and the NS-NS scalar is given by:
 \begin{equation}
 \sigma'= {1\over q'_{E_{p}}}  \Big[ q'_{F_{q}} - {q'_{H}\over \sqrt{2}} \zeta' + {q'_{E_{q}}\over 4} \big( \zeta'^{2}+ 3 \tilde \zeta'^{2} -4 e^{-\phi'}\big ) \Big] \label{sigmaeq}
 \end{equation}
The general solution of the differential equation for the dilaton (\ref{dileqc}) is:
	\begin{equation}
 e^{-\phi'(\tau)}= {q'^2_{E_{q}}+q'^2_{E_{p}}\over 96 c} \sin^{2}\big( 2 \sqrt{3 c}(\tau+\tau_{0}) \big)\label{dilexrr}
 \end{equation}
 where $c$ is given in (\ref{metparb}) and the additional integration constant $\tau_{0}$ can be traded for the value of the dilaton at one boundary. 
 
 \subsection{Mixed NS-R charged case }
In the mixed case all eight charges can be nonzero as long as they satisfy the quadratic and cubic constraints. 
The gravitational constraint (\ref{consta}) or (\ref{constb}) relates  the metric parameter $c$ to the quadratic invariant in the charges.
\begin{equation}\label{metpar}
 c  =  {1\over 96}\Big(2q'_{E} q'_{F} + q'_{E_{q}} q'_{F_{q}}-q'_{E_{p}}q'_{F_{p}} \Big) -{1\over 192} q'^2_{H} - {1\over 64} q'^2_{J}
\end{equation}
The differential equation for $\phi'$ simplifies if expressed in terms of the metric parameter $c$ using (\ref{metpar}).
\begin{equation}\label{dileqa}
(\partial_\tau \phi')^{2}-e^{2\phi'} \;q'^2_{E}-\frac{1}{2}e^{\phi'}\big( q'^2_{E_{q}}+ 4 q'_{E}q'_{J }- q'^2_{E_{p}} \big)
+ 48 c=0
\end{equation}
The cubic constraint in the charges is:
\begin{equation}
q'_{J}\big( q'^2_{J}-q'^2_{H} + 4 q'_{E } q'_{F } \big)+(q'_J-q'_{H})\big(q'_{E_{p}} q'_{F_{q}} - q'_{E_{q}} q'_{F_{p}} \big)+q'_{E}(q'^2_{F_{p}}-q'^2_{F_{q}})+ \ q'_{F}(q'^2_{E_{q}}-q'^2_{E_{p}}) =0
\end{equation}
The R-R-scalars are given by:
\begin{eqnarray}
\zeta'&=&  c_{1}+ c_{2} e^{\phi'} + c_{3} \partial_\tau \phi' \label{zetaxa}
 \\ \tilde\zeta'&=& \tilde c_{1}+ \tilde c_{2} e^{\phi'} + \tilde c_{3} \partial_\tau \phi' \label{zetaxb}
\end{eqnarray}
Where the constants $c_{i},\tilde c_{i}$ are functions of the charges given in the appendix.
The NS-NS scalar $\sigma$ can be obtained from the R-R-scalars and the dilaton:
\begin{equation}
\sigma'= {-q'_{H}+ \frac{ \textstyle q'_{E_{q}}}{\textstyle \sqrt{2}} \zeta' - \frac{\textstyle q'_{E_{p}}}{\textstyle \sqrt{2}}\tilde \zeta' - q'_{E} \zeta'\tilde \zeta' +2 \partial_\tau \phi' \over 2 q'_{E}}
\end{equation}
While  the solution has a complicated dependence on the charges $q_a$, it follows from the dilaton dependence of $\zeta', \tilde \zeta'$  and $ \sigma'$ that the regularity of the solution is determined from (\ref{dileqa}) alone. The solution of this equation is given by:
\begin{equation}
e^{-\phi'(\tau)}= {q'^2_{E_{q}}+ 4 q'_{E}q'_{J} -q'^2_{E_{p}} \over 192 c} \left\{ 1 - \sqrt{ 1 + {768 \;  q'^2_{E} \; c \over (q'^2_{E_{q}}+ 4 q'_{E}q'_{J} -q'^2_{E_{p}})^2 }} \cos [ 4 \sqrt{3 c} (\tau+\tau_{0}) ] \right\} \label{dileqnsr}
\end{equation}

\section{ Supersymmetry and instanton solutions}\label{secsusy}
\setcounter{equation}{0}
In this section, we analyze the condition that preservation of unbroken supersymmetry imposes on the solutions found in the previous section. We will work with the fields prior to analytic continuation in order to derive a condition which is valid for both the pure R-R charged and mixed NS-R charged case.
The transformation rule for the 
gravitino is given by:
\begin{equation}
\delta \psi_{\mu\; \alpha}= (D_{\mu } + Q_{\mu}) \epsilon_{\alpha} + (A_{\mu})_{\alpha}^{\;\; \beta} \epsilon_{\beta}\label{gravsusy}
\end{equation}
Here $D_\mu$ is the covariant derivative and $Q_{\mu}$ is a U(1) K\"{a}hler connection coming from $N=2$ vector multiplets. Since we consider pure $N=2$ supergravity coupled to the universal hypermultiplet, we will set this connection to zero. $A_{\mu}$ is an $Sp(1)$ connection which will be given below.
The hyperino supersymmetry transformation is given by:
\begin{equation}
\delta \xi^{i}= -i V_{\mu}^{\;\;i\alpha} \gamma^{\mu } \epsilon_{\alpha}\label{hypera}
\end{equation}
where $V$ is a quaternionic vielbein. For the universal hypermultiplet, it is given by \cite{Behrndt:1997ch,Strominger:1997eb}
\begin{equation}
(V_{\mu })^{i \alpha}=\left(
\begin{array}{cc}
 u_{\mu}   & v_{\mu} \\
 -\bar v_{\mu}& \bar u_{\mu}
\end{array}
\right)\label{hyperb}
\end{equation}
The $Sp(1)$ connection appearing in the gravitino variation (\ref{gravsusy}) is:
\begin{equation}
(A_{\mu})_{\alpha}^{\;\; \beta}=\left(
\begin{array}{cc}
 {1\over 4} (v_{\mu}-\bar v_{\mu}) &\bar u_{\mu} \\
 - u_{\mu}& -{1\over 4} (v_{\mu}-\bar v_{\mu})
\end{array}
\right)
\end{equation}
where the quantities $u_{\mu}$ and $v_{\mu}$ and their complex conjugates  are:
\begin{eqnarray}
u_{\mu}&=&{1\over 2} e^{-\phi/2}(i \partial_{\mu}\zeta + \partial_{\mu }\tilde\zeta), \quad v_{\mu}={1\over 2}\partial_{\mu}\phi +{i\over 2} e^{-\phi} \Big(\partial_{\mu}\sigma +\tilde \zeta \partial_{\mu}\zeta \Big)\nonumber\\
\bar u_{\mu}&=&{1\over 2} e^{-\phi/2}(-i \partial_{\mu}\zeta +\partial_{\mu }\tilde\zeta), \quad \bar v_{\mu}={1\over 2}\partial_{\mu}\phi -{i\over 2} e^{-\phi}\Big(\partial_{\mu}\sigma +\tilde \zeta \partial_{\mu}\zeta \Big)
\end{eqnarray}
 For $SO(4)$ invariant solutions, the only nontrivial component of the vielbein (\ref{hyperb}) is in the $\tau$ direction. A necessary  condition for an unbroken supersymmetry is that the vielbein $V_{\tau}$ has a zero eigenvalue. This translates into:
 \be
 u_{\tau}\bar u_{\tau}+v_{\tau}\bar v_{\tau}=0 \label{extremb}
 \ee 
 Hence extremality is a necessary condition for supersymmetry.  The eigenvector of $V_\tau$ with zero eigenvalue   can be chosen as:
 \be
\left(
\begin{array}{c}
   \epsilon_{1}\\
    \epsilon_{2}\\
\end{array}
\right) = f(\tau) \left(
\begin{array}{c}
    \bar u_{\tau}\\
   \bar v_{\tau}\\
\end{array}
\right)  \eta_{0}
 \ee
 where $\eta_{0}$ is a $\tau$ independent Weyl spinor and $f(\tau)$ is a function to be determined. With this ansatz the $\delta \psi_{\tau}=0$ component of the gravitino supersymmetry variation becomes:
 
 \begin{eqnarray}
 \dot f  \left(
\begin{array}{c}
    \bar u_{\tau}\\
   \bar v_{\tau}\\
\end{array}
\right) + f  \left(
\begin{array}{c}
    \dot {\bar u}_{\tau}\\
   \dot {\bar v}_{\tau}\\
\end{array}
\right) +  f  \left(
\begin{array}{cc}
 {1\over 4} (v_{\mu}-\bar v_{\mu}) &\bar u_{\mu} \\
 - u_{\mu}& -{1\over 4} (v_{\mu}-\bar v_{\mu})
\end{array}
\right)\  \left(
\begin{array}{c}
    \bar u_{\tau}\\
   \bar v_{\tau}\\
\end{array}
\right) =0\label{integra}
 \end{eqnarray}
 where a dot denotes a derivative with respect to $\tau$.
 There is an integrability condition which is obtained by eliminating $\dot f$ from (\ref{integra}). Using the extremality (\ref{extremb}) the integrability condition can be written as:
\be \dot{\bar u}_{\tau} \bar v_{\tau}- \dot{\bar v}_{\tau} \bar u_{\tau}+{1\over 2} \big(  \bar v_{\tau}-   v_{\tau}\big)  \bar u_{\tau} \bar v_{\tau}=0 \label{integrc}
 \ee
It can be shown for  the pure R-R charged, as well as the NS-R charged solution, the integrability condition is automatically satisfied once the equations of motion are used, without any additional  condition on the charges. Hence any extremal solution for the universal hypermultiplet is BPS and breaks half the supersymmetries.
\subsection{Pure R-R charged solution}
In the extremal limit $c\to 0$ the expression for the dilaton (\ref{dilexrr}) becomes:
\begin{equation}
e^{-\phi(\tau)}= {1\over 8} \big( {q'}_{E_{p}}^{2}+{q'}_{E_{q}}^{2}\big) (\tau+\tau_{0})^{2}
\end{equation}
Note that  $\tau=0$ corresponds to $r=\infty$ and the integration constant $\tau_{0}$ can be related to the value of the dilaton at infinity, i.e., the coupling constant. The action of the instanton is given by the boundary term
\begin{eqnarray}
S_{inst} = \Sigma \nonumber &=& -\frac{1}{\sqrt{2}}\rho_{E_{q}} \big(\zeta'(\infty)- \zeta'(0) \big) - \frac{1}{\sqrt{2}} \rho_{E_{p}} \big(\tilde \zeta'(\infty)- \tilde\zeta'(0) \big)  \\
&=& \sqrt{2({q'}_{E_{p}}^{2} +{q'}_{E_{q}}^{2})}\;  e^{{\phi'}_{\infty}/2}
\end{eqnarray}
With the identification $g_{s}=e^{-\phi'_{\infty}/2}$ the action has the correct behavior to be associated with a D-instanton.

\subsection{Mixed NS-R charged solutions}
In the extremal limit $c\to 0$ the expression for the dilaton (\ref{dileqnsr}) becomes
\begin{equation}
e^{-\phi'(\tau)}= -{2 {q'}_{E}^{2}\over {q'}_{E_{q}}^{2}-{q'}_{E_{p}}^{2}+4 {q'}_{E}{q'}_{J}} + {{q'}_{E_{q}}^{2}-{q'}_{E_{p}}^{2}+4 {q'}_{E}{q'}_{J}
\over 8} (\tau+\tau_{0})^{2}
\end{equation}
As before, the integration constant $\tau_{0}$ can be related to   the dilaton at $r=\infty$. 
The instanton action is given by
 \be
S_{inst} = \Sigma =  {q'}_{E}\big(\sigma'(\infty)-\sigma'(0)\big)-\frac{1}{\sqrt{2}}{q'}_{E_{q}}  \big(\zeta'(\infty)- \zeta'(0) \big)\label{instactnsr}
\ee
While the instanton action above is written in a very compact form, when the formulae (\ref{zetaeqa}) and (\ref{sigmaeq}) are inserted, the result becomes quite complicated in terms of all the charges. It is possible to check that the instanton action is the same as the one proposed by Vandoren et al (equation 3.14 in \cite{deVroome:2006xu}), once the solutions are mapped into each other:
\be \Sigma= \sqrt{4 e^\phi +( \tilde \zeta '(\infty)-\tilde \zeta' (0))^2} \left\{ \left|\frac{q'_{Eq}}{\sqrt{2}}+q'_E \tilde \zeta ' (\infty) \right|+ \frac{q'_E }{2} (\tilde \zeta'(\infty)-\tilde \zeta' (0)) \right\} \ee

\section{Nonsingular wormhole solutions}
 
\setcounter{equation}{0}
The non-extremal solution with $c>0$  can be interpreted as a wormhole solution. The metric factor $e^{U}$ given in (\ref{WHSol}) has two asymptotic regions as $\tau\to 0$ and $\tau\to {\pi \over 4 \sqrt{c}}$, which are connected by a regular wormhole neck.
Since the other hypermultiplet fields depend regularly on the dilaton and its first derivative, the solution can be singular only if the dilaton goes to plus or minus infinity for $\tau \in [0,{\pi \over 4 \sqrt{c}}]$. 

\subsection{Pure R-R charged solution}
The solution (\ref{dilexrr}) implies that the dilaton is nonsingular for $\tau+\tau_0 \in [0, {\pi \over 2  \sqrt{ 3 c}}]$.  Since the radial coordinate ranges from $\tau \in [0,{\pi \over 4 \sqrt{c}}]$ the dilaton can be completely regular between the two asymptotic regions of the wormhole provided that $\tau_0 \leq \frac{2 - \sqrt{3}}{4 \sqrt{3c}} \pi$.\\
Note that the regularity does not depend on the value of the charges as long as the  solution satisfies the quadratic and cubic constraints. This result is in agreement with the bound on the value of the exponential dilaton coupling in the axion-dilaton system  
which was derived in \cite{Giddings:1989bq} (See also \cite{ArkaniHamed:2007js}   for a recent discussion).

\subsection{Mixed NS-R charged solutions}
For the mixed NS-R charged solution the situation is more involved. The expression (\ref{dileqnsr}) for the dilaton is regular as long as
\begin{equation}
 {{q'}_{E_{q}}^{2}+ 4 {q'}_{E}{q'}_{J} -{q'}_{E_{p}}^{2}\over      \sqrt{({{q'}'}_{E_{q}}^{2}+ 4 {q'}_{E}{q'}_{J} -{q'}_{E_{p}}^{2})^{2}+768\;  {q'}_{E}^{2} \; c }}\geq  \cos [ 4 \sqrt{3 c} (\tau+\tau_{0})]
\end{equation}
This inequality constraints the maximal range of the radial coordinate $\tau$.  The wormhole solution will therefore be regular if this range fits into the interval  $\tau \in [0,{\pi \over 4 \sqrt{c}}]$.
There can be regular solutions for some value of the integration constant $\tau_0$ only if:
 \begin{equation}
({q'}_{E_{q}}^{2}+ 4 {q'}_{E}{q'}_{J} -{q'}_{E_{p}}^{2}) ^{2} \geq  \cos^2\left( {\frac{\sqrt{3}}{2} \pi } \right)  \Big[ ({q'}_{E_{q}}^{2}+ 4 {q'}_{E}{q'}_{J} -{q'}_{E_{p}}^{2}) ^{2} + 768 \;  {q'}_{E}^{2} \; c \Big]\label{nschcon}
\end{equation}
 
Hence, there are regular wormhole solutions which carry both NS and R charges. Note however that if the NS charge ${q'}_{E}$ is increased, the condition (\ref{nschcon}) 
will eventually be violated. In particular, the special case of a purely NS charged wormhole (with ${q'}_{E_{q}}=0, {q'}_{E_{p}}=0$) violates the bound. This is in agreement with the condition on the dilaton coupling of dilaton-axion wormholes \cite{Giddings:1989bq}, since the coupling  of the dilaton to the NS-NS axion $\sigma'$ in the universal hypermultiplet action (\ref{unione})  is outside the bound for regular wormhole solutions. 

\section{Conclusions}


Instanton and wormhole solutions for the universal hypermultiplet were constructed as saddle points of the Euclidean path integral representing transition amplitudes between states with definite axion shift charge.  This approach provides us with a physically intuitive explanation for the flip of sign in the kinetic terms of axionic scalars in the Euclidean action. Furthermore, it explains why there are two different analytic continuations for the R-R charged and NS-R charged solutions:   they correspond to different representations of the Heisenberg algebra of shift charges in  the Stone-von Neumann theorem. 

The general solutions were constructed employing the global $SU(2,1)$ symmetry of the universal hypermultiplet action. In the extremal limit, the solutions found are equivalent to the ones obtained in the literature using different methods for the analytic continuation. 

For the universal hypermultiplet, extremal instanton solutions are always BPS and no non-BPS extremal instantons exists.  We found some new regular wormhole solutions carrying both NS and R-R charges, in contrast to the purely NS charged wormhole which is singular.

\bigskip

\noindent{\bf Acknowledgements}

\medskip

\noindent This work is supported in part by NSF grants PHY-0456200 and PHY-0757702. MG is grateful to the Department of Physics and Astronomy, Johns Hopkins University and the International Centre for Theoretical Science, TIFR, Mumbai, for hospitality while this paper was completed.

\newpage

\appendix
\section{Details on the mixed NS-R charged solution}
\setcounter{equation}{0}
In this appendix we give some details on the general NS-R charged solution presented in section 5.2. The constants $c_{i}$ in the expression for $\zeta$ in (\ref{zetaxa}) are given by
 \begin{eqnarray}
c_{1}&=&  {1 \over \sqrt{2} D}  \Big[ (q'^2_{E_q}-q'^2_{E_p} ) (q'_{E_q} q'_{H} + q'_{E_p} q'_{J})+ 8 q'^2_{E} ( q'_{F_p} q'_{J}-q'_{E_p} q'_{F} ) + 2 q'_{E} q'^2_{E_p} q'_{F_p} \nonumber \\ 
&& \quad + 2 q'_E q'_{E_q} (q'_{E_q} q'_{F_p} + 2 q'_{H} q'_{J}) +  
    2q'_E q'_{E_p} ( q'^2_{H} + 5 q'^2_{J}-2 q'_{E_q} q'_{F_q} )\Big]  \\
  c_{2}&=&{\sqrt{2} q'_{E}\over D} \Big[  - q'^3_{E_p} + 2 q'_{E} (2 q'_{E} q'_{F_p} + q'_{E_q} q'_{H}) + q'_{E_p} ( q'^2_{E_q} + 6 q'_{E} q'_{J})\Big]
 \\
  c_{3} &=& {\sqrt{2} \over D} \Big(  q'^2_{E_p} q'_{E_q} -  q'^3_{E_q} - 4 q'^2_{E} q'_{F_q} - 2 q'_{E} q'_{E_p} q'_{H} - 6 q'_{E} q'_{E_q} q'_{J}\Big) 
 \end{eqnarray}
%
%
%
The constants $\tilde c_{i}$ in the expression for $\tilde\zeta$ in (\ref{zetaxa}) are given by
\begin{eqnarray}
\tilde{c}_{1} &=& {1 \over \sqrt{2} D} \Big[ (q'^2_{E_q}-q'^2_{E_p} ) (q'_{E_p} q'_{H} + q'_{E_q} q'_{J}) + 8 q'^2_{E} (q'_{F_q} q'_{J}-q'_{E_q} q'_{F}) - 2 q'_{E} q'^2_{E_p} q'_{F_q} \nonumber \\
&& \quad +4 q'_E q'_{E_p} (q'_{E_q} q'_{F_p} + q'_{H} q'_{J}) +  2 q'_E q'_{E_q} (q'^2_{H} + 5 q'^2_{J}-q'_{E_q} q'_{F_q}) \Big]  \\
\tilde{c}_{2} &=&   {\sqrt{2} q'_{E}\over D}  \Big(- q'^2_{E_p} q'_{E_q} + q'^3_{E_q} + 4 q'^2_{E} q'_{F_q} + 2 q'_{E} q'_{F_p} q'_{H} + 6 q'_{E} q'_{E_q} q'_{J}\Big) \\
\tilde{c}_{3} &=&  {\sqrt{2} \over D} \Big[ q'^3_{E_p} - 2 q'_{E} (2 q'_{E} q'_{F_p} + q'_{E_q} q'_{H}) - q'_{E_p} ( q'^2_{E_q} + 6 q'_{E} q'_{J})\Big]
\end{eqnarray}
%
%
%
All constants have the same denominator which is given by
\be
D=\frac{ (q'^2_{E_p} - q'^2_{E_q})^2}{2} + 4 q'^2_{E} ( q'_{E_q} q'_{F_q}-q'_{E_p} q'_{F_p}) + 4 ( q'^2_{E_q}-q'^2_{E_p}) q'_E q'_{J} + 2 q'^2_{E} (4 q'_{E} q'_{F} - q'^2_{H} + q'^2_{J}) 
\ee

\section{Relation of universal hypermultiplet and double tensor multiplet intanton solutions}
\setcounter{equation}{0}
An alternative formulation of the universal hypermultiplet is given dualizing the axions $\sigma'$ and $\zeta'$ into two rank three antisymmetric tensor field strengths  $H^{(1)}_{\mu\nu\rho}, H^{(2)}_{\mu\nu\rho}$. The action is given by
\begin{eqnarray}
S&=& \int d^{4}x \sqrt{g} \;\Big\{ {1\over 2} (\partial_{\mu} \phi')^{2}+ {1\over 2}e^{-\phi'} (\partial_{\mu} \tilde \zeta')^{2} +{1\over 2} e^{\phi'}   H^{(1)}_{\mu\nu\rho}H^{(1)\; \mu\nu\rho} - e^{\phi'}  \tilde\zeta'  H^{(1)}_{\mu\nu\rho}H^{(2)\; \mu\nu\rho}   \nonumber \\
&& \quad  +{1\over 2} \big( e^{2\phi'}+  e^{\phi'}  \tilde\zeta'^{2}\big)   H^{(2)}_{\mu\nu\rho}H^{(2)\; \mu\nu\rho} \Big\}
\label{unitwo}
\end{eqnarray}
In \cite{Theis:2002er,deVroome:2006xu} instanton solutions which carry both $H^{(1)}$ and $H^{(2)}$ charges were constructed, in this appendix we make the relation of these solutions with the mixed NS-R charged instanton solutions explicit.
The solution of \cite{Theis:2002er,deVroome:2006xu} is parameterized in terms of two harmonic functions $h$ and $p$ 
\be
h(t)=h_{0}+ Q_{h} t, \quad  p(t)=p_{0}+ Q_{p} t,
\ee
and is given by

\begin{eqnarray}
\phi' &=& - \log\left[ ({h^{2}-p^{2} )/ 4}\right]  \label{solvana}\\
\tilde \zeta' &=& \tilde \zeta'_{0} - 4 {p\over h^2 - p^2} \\
\zeta'&=&  \zeta'_{0} - 4 {h\over h^2 - p^2} \\  
 \sigma' &=&  \sigma_{0}' + 4 h { (h^{2}-p^{2} )\tilde\zeta'_{0}  -2 p  \over  \big(  h^{2}-p^{2}\big)^{2}}   \label{solvand}
\end{eqnarray}
$\zeta_{0}$ and $\sigma_{0}$ are integration constants coming from the dualization of the tensor fields to the scalars $\zeta$ and $\sigma$.  Using this parameterization it is easy to show that the instanton action (\ref{instactnsr}) agrees with the one found in \cite{deVroome:2006xu}.  

\section{Conserved charges after analytic continuation}
\setcounter{equation}{0}
In this appendix we give details on the conserved currents after the analytic continuations. For pure R-R solutions the analytic continuation is given in (\ref{anaconrr}), the primed currents become:
\bea
j'_{E_{p}\, \mu}& = & - \sqrt{2} e^{\phi'} \partial_{\mu }\tilde \zeta'  \\
j'_{E_{q}\, \mu}& = & - \sqrt{2} e^{\phi'} \partial_{\mu }\zeta'   \eea
The other conserved currents are:
\bea j'_{H\; \mu} &=& \frac{1}{\sqrt{2}} \tilde\zeta' j'_{E_p \; \mu} + \frac{1}{\sqrt{2}} \zeta' j'_{E_q \; \mu}  + 2 \partial_{\mu} \phi' \\
j'_{J \;\mu} &=&  \frac{1}{\sqrt{2}} \zeta' j'_{E_p \; \mu} - \frac{1}{\sqrt{2}} \tilde \zeta' j'_{E_q \; \mu}  \eea
\bea
j'_{F_p \mu} &=&  \left( e^{\phi'} + \frac{\tilde \zeta'^2}{4}-\frac{3 \zeta'^2}{4} \right)j'_{E_p \mu} + \left( \frac{3\tilde \zeta' \zeta'}{2} - \sigma'  \right) j'_{E_q \mu} + \sqrt{2} \tilde \zeta' \partial_\mu \phi' \\
j'_{F_q \mu}  &=&   \left( \sigma' + \frac{ \tilde \zeta' \zeta'}{2}  \right) j'_{E_p \mu}+ \left( e^{\phi'}+  \frac{\zeta'^2}{4} - \frac{3 \tilde \zeta'^2}{4} \right) j'_{E_q  \mu} + \sqrt{2} \zeta' \partial_\mu \phi' \\
 j'_{F \mu}  &=&   \frac{1}{\sqrt{2}} \left(\textstyle e^{\phi'} \zeta' - \frac{\zeta'^3}{4} + \frac{ \tilde \zeta'^2 \zeta'}{4} - \sigma' \tilde \zeta' \right)j'_{E_p \mu} + \frac{1}{\sqrt{2}}\left( -\textstyle e^{\phi'} \tilde \zeta'  + \frac{\tilde \zeta'^3}{4} + \frac{3\tilde \zeta' \zeta'^2}{4} - \sigma' \zeta'  \right) j'_{E_q \mu} \quad \\
&&   + \left(\textstyle \tilde \zeta' \zeta' -2 \sigma  \right) \partial_\mu \phi' 
\eea
For mixed NS-R solutions the analytic continuation is given in (\ref{anaconnsr}), the shift currents are: 
\bea
j'_{E_{p}\, \mu}& = & - \sqrt{2} [e^{\phi'} \partial_{\mu }\tilde \zeta'   + e^{2\phi'}  \zeta' \big( \partial_{\mu}\sigma'  + \tilde \zeta' \partial _{\mu }\zeta' \big)] \\
j'_{E_{q}\, \mu}& = & - \sqrt{2} [e^{\phi'} \partial_{\mu }\zeta'   + e^{2\phi'}  \tilde \zeta' \big( \partial_{\mu}\sigma'  + \tilde \zeta' \partial _{\mu }\zeta' \big)] \\
j'_{E\, \mu} & = & e^{2 \phi'} \big( \partial_\mu \sigma' + \tilde\zeta' \partial_{\mu} \zeta' \big) 
\eea
The other conserved currents are:
\bea && j'_{H\; \mu} = - \frac{1}{\sqrt{2}} \tilde\zeta' j'_{E_p \; \mu} + \frac{1}{\sqrt{2}} \zeta' j'_{E_q \; \mu} - (2 \sigma' + \tilde \zeta' \zeta') j'_{E \; \mu} + 2 \partial_{\mu} \phi' \\
&& j'_{J \;\mu} = - \frac{1}{\sqrt{2}} \zeta' j'_{E_p \; \mu} + \frac{1}{\sqrt{2}} \tilde \zeta' j'_{E_q \; \mu} + \frac{\tilde \zeta'^2-\zeta'^2}{2} j'_{E \mu} \\
&&j'_{F_p \mu} =  \textstyle \left( \textstyle e^{\phi'}- \frac{\tilde \zeta'^2}{4}-\frac{3 \zeta'^2}{4} \right)j'_{E_p \mu} + \left( \textstyle \frac{3\tilde \zeta' \zeta'}{2} + \sigma'  \right) j'_{E_q \mu}+ \sqrt{2} \zeta' \left( \textstyle e^{\phi'} + \frac{\tilde \zeta'^2 - \zeta'^2}{2}  \right) j'_{E \mu} \nonumber \\ &&\quad \quad \quad + \sqrt{2} \tilde \zeta' \partial_\mu \phi' \\
&&\textstyle j'_{F_q \mu}  =   \left( \textstyle \sigma' - \frac{ \tilde \zeta' \zeta'}{2}  \right) j'_{E_p \mu}+ \left(\textstyle e^{\phi'}+  \frac{\zeta'^2}{4} + \frac{3 \tilde \zeta'^2}{4} \right) j'_{E_q  \mu} + \sqrt{2} \tilde \zeta' \left( \textstyle e^{\phi'} + \frac{\tilde \zeta'^2 -\zeta'^2}{2} \right) j'_{E \mu} \nonumber \\&&
\quad \quad \quad+ \sqrt{2} \zeta' \partial_\mu \phi' \\
&&j'_{F \mu}  =   \frac{1}{\sqrt{2}} \left(\textstyle - e^{\phi'} \zeta' + \frac{\zeta'^3}{4} + \frac{ \tilde \zeta'^2 \zeta'}{4} + \sigma' \tilde \zeta' \right)j'_{E_p \mu} + \frac{1}{\sqrt{2}}\left( \textstyle e^{\phi'} \tilde \zeta'  + \frac{\tilde \zeta'^3}{4} - \frac{3\tilde \zeta' \zeta'^2}{4} - \sigma' \zeta'  \right) j'_{E_q \mu}+ \quad \nonumber  \\
&& \quad  \left( \textstyle e^{2 \phi'} + \frac{3\tilde \zeta'^4 + 3\zeta'^4}{16} + \sigma'^2 + \tilde \zeta' \zeta' \sigma' + e^{\phi'}( \tilde \zeta'^2 - \zeta'^2) - \frac{\tilde \zeta'^2 \zeta'^2}{8}  \right) j'_{E \; \mu} - \left(\textstyle 2 \sigma' + \tilde \zeta' \zeta'  \right) \partial_\mu \phi'\quad 
 \eea

\end{document}